\documentclass [6pt,a4paper]{article}
\usepackage{bbm}
\usepackage{amsfonts}
\usepackage{mathrsfs}
\usepackage {amssymb}
\usepackage {amsmath}
\usepackage{amsthm}
\usepackage{latexsym}
\usepackage{booktabs}
\def\dse#1{\vskip 0.6cm\noindent
        {\large\bf #1}
        \vskip 0.4cm}

\def\dec#1{\vskip 0.6cm\noindent
        {\large\bf #1}
        \vskip 0.4cm}
\usepackage{lastpage}
\usepackage{epsfig}

\begin{document}
\begin{center}
\textbf{\large{The symbol-pair distance distribution of repeated-root cyclic codes over $\mathbb{F}_{p^m}$ }}\footnote { E-mail addresses:
 zhushixin@hfut.edu.cn(S.Zhu), sunzhonghuas@163.com(Z.Sun),liqiwangg@163.com(L.Wang)
}
\end{center}

\begin{center}
{ { Shixin Zhu, \  Zhonghua Sun, \ Liqi Wang } }
\end{center}

\begin{center}
\textit{School of Mathematics, Hefei University of
Technology, Hefei 230009, Anhui, P.R.China 
}
\end{center}

\noindent\textbf{Abstract:} Symbol-pair codes are proposed to protect against pair errors in symbol-pair read channels. One of the most important task in symbol-pair coding theory  is to determine the minimum pair-distance of symbol-pair codes. In this paper, we investigate the symbol-pair distances of cyclic codes of length $p^e$ over $\mathbb{F}_{p^m}$. The exact symbol-pair distances of all cyclic codes of such length are determined.\\


\dec{1~~Introduction}
Recently, Cassuto and Blaum [1,2] proposed the model of symbol-pair read channels. Such channels are mainly motivated by storage technologies, due to physical limitations, each channel read contains contributions from two adjacent symbols. For this new channels, the codes defined as usual over some discrete symbol alphabet, but whose reading from the channel is performed as overlapping pairs of symbols. More precisely, let $\Xi$ be the alphabet consisting of $q$ elements, the code is defined as a subset $\mathcal{C}\subset \Xi^n$, the decoder pair-dec : $ (\Xi,\Xi)^n \rightarrow \mathcal{C}$. As usual, a pair-error is define as a pair-read in which one or more of the symbols are read in error. They designed  symbol-pair codes to protect against pair-errors in symbol-pair read channels. This work has laid out a coding theoretic framework to combat pair-errors over symbol-pair channels.

Cassuto and Blaum [1,2] defined the pair-distance as the Hamming distance over the alphabet $(\Xi,\Xi)$ for achieving correctability of symbol-pair errors, as well as code construction, decoding methods, and asymptotic bounds. Later, Cassuto and Litsyn [3] (also see [2]) using cyclic codes to construct symbol-pair codes. Lower bounds on pair-distance of cyclic codes are obtained by using the discrete Fourier transform and BCH bounds, [3, Theorem 4, or, 2, Theorem 10]. It shown that if  simple-root $[n,k, d_H]$ cyclic codes has at least $d_H$ roots, then the minimum pair-distance $d_p\geq d_H+2$. For prime length $n$, by using the Hartmann-Tzeng Bound, this lower bound can improved to $d_H+3$ for some constraint condition. In the following papers, Kai et al. [10] showed that [3, Theorem 4, or, 2, Theorem 10] can be generalized to simple-root constacyclic codes case. Recently, this two lower bounds of minimum pair-distance of simple-root cyclic (constacyclic) codes are generalized to repeated-root cyclic (constacyclic) codes by Chen et al. [7]. For binary cyclic codes, Yaakobi et al. [11, Theorem 4, also, 12, Theorem 1] given an improved lower bound on the minimum pair-distance for cyclic codes: If $\mathcal{C}$ is an $[n,k,d_H]$ cyclic codes ($k>1$), then $d_p\geq d_H+ \lceil \frac{d_H}{2} \rceil$. Unfortunately, little work has been done on the exact values of the symbol-pair distances of cyclic codes. Only a few of the pair-distances of symbol-pair codes are known. Similar to the Hamming distance, Chee et al. [5, 6] established a Singleton-type bound for symbol-pair codes and constructed infinite families of maximum distance separable (MDS) symbol-pair codes. Later, Kai et al. [10] constructed some MDS symbol-pair codes via the constacyclic codes with minimum pair-distance five and six. Recently, Chen et al. [7] obtained some MDS symbol-pair codes with minimum pair-distance seven and eight through the cyclic codes of length $3p$ over $\mathbb{F}_p$. By using [8, Theorem 4.11], we know that cyclic codes of length $p$ over $\mathbb{F}_{p^a}$ are MDS codes, so their are also MDS symbol-pair codes. This inspires us to study the minimum pair-distances of repeated-root cyclic codes. For the class of cyclic codes of length $p^s$ over $\mathbb{F}_{p^a}$, their Hamming distance had been completely determined in [8]. This will help us to determine the exact values of their pair-distances.

In this paper, we study the minimum pair-distance of cyclic codes of length $p^e$ over $\mathbb{F}_{p^m}$. In Section 2, we give some background and recall some basic results of the pair-distance and  cyclic codes. In Section 3, we compute the pair-distance of all cyclic codes of length $p^e$ over $\mathbb{F}_{p^m}$. Finally, the conclusion of paper is reached in section 4.

\dec{2~~Preliminaries}
Let $\Xi$ be the alphabet consisting of $q$ elements. Each element in $\Xi$ is called a symbol. We use $\Xi^n$ to denote the set of all $n-$tuples, where $n$ is a positive integer. Let $\textbf{x}=(x_0,x_1,\cdots,x_{n-1})\in \Xi^n$, the symbol-pair read vector of $\textbf{x}$ is defined as
\begin{align*}
\pi(\textbf{x})=[(x_0,x_1),(x_1,x_2),\cdots,(x_{n-2},x_{n-1}),(x_{n-1},x_0)].
\end{align*}
Every vector $\textbf{x}\in \Xi^n$ has an unique pair representation $\pi(\textbf{x})\in ( \Xi \times \Xi)^n$. For any two symbol pairs $(a,b)$ and $(f,g)$, say $(a,b)=(f,g)$ if both $a=c$ and $b=g$. Recall that the Hamming weight of a vector $\textbf{x}$ is defined by $\omega_{H}(\textbf{x})=|\{ i\in \mathbb{Z}_n|x_i\neq0\}|$, where $\mathbb{Z}_n$ denotes the ring $\mathbb{Z}/n\mathbb{Z}$. Define the pair-weight of a vector $\textbf{x}$ is $\omega_p(\textbf{x})=|\{ i\in \mathbb{Z}_n| (x_i,x_{i+1})\neq (0,0)\}|$. For two vectors $\textbf{x},~\textbf{y} \in \Xi^n$, the pair-distance $\textbf{x}$ and $\textbf{y}$ is define as $d_p(\textbf{x},\textbf{y})=|\{ i\in \mathbb{Z}_n| (x_i,x_{i+1})\neq (y_i,y_{i+1})\}|$.

Define a code $\mathcal{C}\subseteq \Xi^n$, and let $d_p(\mathcal{C})=min \{d_p(\textbf{x},\textbf{y})|\textbf{x},\textbf{y}\in \mathcal{C},\textbf{x}\neq \textbf{y}\}$ be the minimum pair-distance of $\mathcal{C}$. Similar to the classical case, a code over $\Xi$ of length $n$ with size $M$ and minimum pair-distance $d$ is called an $(n,M,d)$ symbol-pair code. If $\mathcal{C}$ is a linear code, then the minimum pair-distance of $\mathcal{C}$ is the smallest pair-weight of nonzero codewords of $\mathcal{C}$. The minimum pair-distance is one of the important parameters of symbol-pair codes. A code $\mathcal{C}$ with minimum pair-distance $d$ can correct $t$ pair-errors if and only if $d\geq 2t+1$ [1 and 2, Proposition 3]. This distance distribution is very difficult to compute in general, however, for the class of cyclic codes of length $p^s$ over $\mathbb{F}_{p^m}$, their Hamming distance has been completely determined in [8]. This will help us to determine the exact values of their pair-distances.

The structure of cyclic code of length $p^s$ over $\mathbb{F}_{p^m}$ has been extensively studied in [8], which is \\

\noindent \textbf{Proposition 2.1.} ([8], Theorem 6.4) \emph{Let $\mathcal{C}$ be a cyclic code of length $p^e$ over $\mathbb{F}_{p^m}$, then $\mathcal{C}=\langle (x-1)^i\rangle\subseteq \frac {\mathbb{F}_{p^m}[x]}{\langle x^{p^e}-1\rangle}$, for $i\in\{0,1,\cdots,p^e\}$. The Hamming distance $d_H$ of $\mathcal{C}$ is determined by
\begin{equation*}d_H=
\begin{cases}1& if~i=0,\\
\beta+2 & if~\beta p^{e-1}+1\leq i\leq (\beta +1)p^{e-1}~where~0\leq \beta \leq p-2,\\
(t+1)p^k& if~p^e-p^{e-k}+(t-1)p^{e-k-1}+1\leq i \leq p^e-p^{e-k}+tp^{e-k-1},\\
& where~1\leq t \leq p-1,~and~1\leq k \leq e-1,\\
0 & if~i=p^e.
\end{cases}
\end{equation*}}\\

Throughout this paper, $d_H$ and $d_p$ denote the usual Hamming distance and the pair-distance, respectively. Cassuto and Blaum [2, Theorem 2] showed the following relationship between the pair-distance and the Hamming distance.\\

\noindent \textbf{Proposition 2.2.} ([2], Theorem 2) \emph{For any $\textbf{x},\textbf{y}\in \Xi^n$ with $0< d_H(\textbf{x},\textbf{y})<n$, define the set $S_H=\{j\mid x_j\neq y_j \}$. Let $S_H=\bigcup_{l=1}^LB_l$ be a minimal partition of the set $S_H$ to subsets of consecutive indices ( Each subset $B_l=[s_l,e_l]$ is the sequence of all indices between $s_l$ and $e_l$, inclusive, and $L$ is the smallest integer that achieves such partition ). Then
\begin{align*}
d_p(\textbf{x},\textbf{y})= d_H(\textbf{x},\textbf{y})+L.
\end{align*}}\\

The minimum pair-distance $d_p$ is an important parameter in determining the error-correcting capability of $\mathcal{C}$. Thus it is significant to find symbol-pair codes of fixed length $n$ with pair-distance $d$ as large as possible. In [5] and [6], the authors have proved the following Singleton bound.\\

\noindent \textbf{Proposition 2.3.} (Singleton Bound) \emph{Let $p^m \geq 2$ and $2\leq d_p \leq n$, If $\mathcal{C}$ is an $(n,M,d_p)$ symbol-pair code, then $M\leq p^{m(n-d_p+2)}$.}\\

A symbol-pair code achieving the Singleton bound is a maximum distance separable (MDS) symbol-pair code.

\dse{3~~Pair-distance of cyclic codes}
In this section, we will determine the pair-distance of cyclic codes of length $p^e$ over $\mathbb{F}_{p^m}$. Firstly, for two codes $\mathcal{C}_1, \mathcal{C}_2\subseteq \mathbb{F}_{p^m}^{p^e}$ with $\mathcal{C}_1 \subseteq \mathcal{C}_2$, we have $d_p(\mathcal{C}_1)\geq d_p(\mathcal{C}_2)$. Secondly, for $i=0,1,\cdots,p^e$, we denote each code $\mathscr{C}_i=\langle (x-1)^i\rangle$. Obvious, $d_p(\mathscr{C}_0)=2$, and we define $d_p(\mathscr{C}_{p^e})=0$. If $p^e=2$, then $d_p(\mathscr{C}_1)=2$. If $e=1$, we have the following results.\\

\noindent \textbf{Proposition 3.1.}\emph{ If $p$ is prime and $e=1$, then
\begin{align*}d_p(\mathscr{C}_i)=
\begin{cases}0& if~i=p,\\
i+2& if ~0\leq i\leq p-2,\\
p& if~i=p-1.
\end{cases}
\end{align*}}\\

\textbf{Proof.} If $e=1$, from Proposition 2.1, we have $d_H(\mathscr{C}_i)=i+1$, where $i\in\{0,1,\cdots,p-2\}$. From Proposition 2.2, we have $d_p(\mathscr{C}_i)\geq i+2$. On the other hand, we have $\omega_p((x-1)^i)=i+2$, for $0\leq i\leq p-2$. So $d_p(\mathscr{C}_i)=i+2$. If $i=p-1$, it is easy check that $d_H(\mathscr{C}_{p-1})=p $. \qed\\

From Proposition 3.1, we have $\mathscr{C}_i$ is a MDS symbol-pair code over $\mathbb{F}_{p^m}$, where $i\in \{0,1,\cdots,p-2\}$. We will always assume $e\geq 2$ in the following paper.\\

\noindent \textbf{Proposition 3.2.} \emph{If $e\geq 2$, for $0\leq i\leq p^{e-1}$, then
\begin{align*}d_p(\mathscr{C}_i)=
\begin{cases}0& if~i=p^e,\\
2& if~i=0,\\
3& if~i=1,\\
4& if ~2\leq i\leq p^{e-1}.
\end{cases}
\end{align*}}

\textbf{Proof.} If $e\geq 2$, then $p^e\geq 4$, since then we have $\omega_p(x-1)=3$, so $d_p(\mathscr{C}_1)\leq3$. From Proposition 2.2, we have $d_p(\mathscr{C}_1)\geq d_H(\mathscr{C}_1)+1=3$. Hence, $d_p(\mathscr{C}_1)=3$.

Firstly, $d_p(\mathscr{C}_2)\leq4$ since $\omega_p(x^p-1)=4$, and $d_p(\mathscr{C}_2)\geq d_p(\mathscr{C}_1)=3$. From Proposition 2.2, if $\omega_p(c(x))=3$ if and only if $c(x)=x^i(a+bx)$ where $a,b\in \mathbb{F}_{p^m}^*$ and $i$ is some integer. Hence, $d_p(\mathscr{C}_2)\neq3$. So, $d_p(\mathscr{C}_2)=4$. Secondly, $d_p(\mathscr{C}_{p^{e-1}})\leq \omega_p(x^{p^{e-1}}-1)=4$. So, for $2\leq i\leq p^{e-1}$, $4=d_p(\mathscr{C}_2)\leq d_p(\mathscr{C}_i)\leq d_p(\mathscr{C}_{p^{e-1}})\leq 4$, i.e. $d_p(\mathscr{C}_i)=4$.\qed\\

\noindent \textbf{Proposition 3.3.} \emph{For $0\leq i\leq p-1$, then \begin{align*}d_p(\mathscr{C}_{p^e-p+i})=
\begin{cases}(i+2)p^{e-1}& if ~0\leq i\leq p-2,\\
p^e& if~i=p-1.
\end{cases}
\end{align*}}\\

\textbf{Proof.} If $i = p-1$, $(x-1)^{p^e-1}=\sum_{j=0}^{p^e-1}x^j$, for any $c(x)\in \mathscr{C}_{p^e-1}$, we have $c(x)=a(x-1)^{p^e-1}$, where $a\in \mathbb{F}_{p^m}$. Hence, $d_p(\mathscr{C}_{p^e-1})=p^e$.

For any fixed $0\leq i\leq p-2$, let $c(x)$ be any nonzero element of $\mathscr{C}_{p^e-p+i}$, then there is a nonzero element $f(x)$ and $deg(f(x))<p-i$ such that $c(x)=(x-1)^{p^e-p}(x-1)^if(x)$. Denote $g(x)=(x-1)^if(x)$, from Proposition 3.1, we have $\omega_p(g(x))\geq i+2$, and

If $deg(g(x))<p-1$ or $g(0)=0$, then $\omega_p(c(x))=p^{e-1}\omega_p(g(x))\geq (i+2)p^{e-1}$.

If $deg(g(x))=p-1$ and $g(0)\neq 0$, then $\omega_p(c(x))=p^{e-1}(\omega_p(g(x))-1)$. From Proposition 2.1, we have $\omega_H(g(x))\geq i+1$. If $\omega_H(g(x))\geq i+2$, then $\omega_p(g(x))\geq \omega_H(g(x))+1\geq i+3$, then $\omega_p(c(x))\geq (i+2)p^{e-1}$. If $\omega_H(g(x))=i+1$, note that $deg(g(x))=p-1$ and $g(0)\neq 0$, but $i+1\leq p-1$, then $\omega_p(g(x))\geq i+3$. Hence, $\omega_p(\mathscr{C}_{p^e-p+i})\geq (i+2)p^{e-1}$. On the other hand, $\omega_p((x-1)^{p^e-p+i})=(i+2)p^{e-1}$. So, we have $d_p(\mathscr{C}_{p^e-p+i})=(i+2)p^{e-1}$. \qed\\

\noindent \textbf{Proposition 3.4.} \emph{Let $\beta,e $ be integer such that $1\leq \beta\leq p-2$ and $e\geq 2$, then $d_p(\mathscr{C}_{\beta p^{e-1}+1})=2(\beta+2)$.}\\

\textbf{Proof.} For any fixed $1\leq \beta \leq p-2$, let $c(x)$ be any nonzero element of $\mathscr{C}_{\beta p^{e-1}+1}$, then there is a nonzero element $f(x)$ and $deg(f(x))<(p-\beta)p^{e-1}-1$ such that $c(x)=(x-1)^{\beta p^{e-1}}(x-1)f(x)$. Denote $g(x)=(x-1)f(x)$, then $\omega_H(g(x))\geq 2$, and
\begin{align*}
c(x)=\sum_{j=0}^\beta (-1)^{\beta-j}{\beta \choose j}x^{jp^{e-1}}g(x).
\end{align*}

Obviously, the cyclic shift of $c(x)$ have the same symbol-pair weight. Hence, we will consider $g(0)\neq 0$.

If $\omega_H(c(x))= 2\beta +3$, we will shown that $d_p(c(x))>2\beta +4$. Indeed, if $d_p(c(x))=2\beta +4$, applying the cyclic shift a certain number of times on $c(x)$ if necessary, then $c(x)$ have the form $c_0+c_1x+\cdots+c_{2\beta+2}x^{2\beta+2}$, but $2\beta+2<\beta p^{e-1}+1$ except for $e=2$ and $p=3$. If $e=2$ and $p=3$, then $(x-1)^{\beta p^{e-1}+1}=(x-1)^4=x^4+2x^3+2x+1$, from this we have $c_0-c_4+(c_1+c_4)x+c_2x^2+(c_3+c_4)x^3\in \langle (x-1)^4\rangle$, this is contradictory. Hence, if $\omega_H(c(x))\geq 2\beta +3$, we have $\omega_p(c(x))\geq 2\beta +5$.

\textbf{Case 1.} If there is no nonzero term $g_ix^i$ of $g(x)$ such that $i\equiv0~mod~p^{e-1}$ ($i>0$), then $c(x)$ have the form $a((-1)^\beta\star\cdots\star a_1 \star\cdots\star a_2 \star\cdots \star a_{\beta-1} \star\cdots\star a_\beta \star \cdots)$, where $a_i=(-1)^{\beta-i}{\beta \choose i}~mod~p$, the star is some elements of $\mathbb{F}_{p^m}$, $a\in \mathbb{F}_{p^m}^\ast$. Next, we will shown that $d_p(c(x))\geq 2\beta +4$.

Obvious, $\omega_H(c(x))\geq \beta+3$. If $\omega_H(c(x))= \beta+3$, then $c(x)=(x-1)^{\beta p^{e-1}}+bx^i(x^j-1)$, for some $b \in \mathbb{F}_{p^m}^*$. From $(x-1)^{\beta p^{e-1}+1}\mid c(x)$, we deduced that $(x-1)^{\beta p^{e-1}}\mid ax^i(x^j-1)$, i.e. $(x-1)^{\beta p^{e-1}}\mid (x^j-1)$, from this, we have $\beta=1$ and $j=j'p^{e-1}$, $gcd(j',p)=1$. So, $\omega_p(c(x))\geq 2+5=2\beta+5$ except for $\omega_p(a(x-1)^{p^{e-1}+1})=6=2(\beta+2)$, where $a\in \mathbb{F}_{p^m}^\ast$.

If $\beta+4\leq\omega_H(c(x))\leq 2\beta+2$, this means the star have $\omega_H(c(x))-\beta-1$ number of nonzero, if $\omega_H(c(x))-\beta-1<p^{e-1}-1$, we can deduced, $\omega_p(c(x))\geq \omega_H(c(x))+\beta+1 \geq 2\beta+5$. If $e=2$, $\beta=p-2$ and $\omega_H(c(x))=2(p-1)$, then $\omega_p(c(x))\geq \omega_H(c(x))+\beta=3p-4\geq 2(p-2)+5$, for $p\geq 5$. If $p=3$ and $e=2$, then $\omega_H(c(x))=4$, this contradiction with $\omega_H(c(x))\geq \beta+4=5$. Hence, we have $\beta+4\leq\omega_H(c(x))\leq 2\beta+2$, then $\omega_p(c(x))\geq 2\beta+5$.

\textbf{Case 2.} If there are not less than one nonzero term $g_ix^i$ of $g(x)$ such that $i \equiv0~mod~p^{e-1}$. Let $g_1(x)=g_{i_1}x^{i_1p^{e-1}}+\cdots+g_{i_s}x^{i_sp^{e-1}}$ and $g_2(x)=g(x)-g_1(x)$, where $0\leq i_1<i_2<\cdots<i_s\leq p-\beta-1$. Then
\begin{align*}
\omega_H(c(x))=\omega_H((x-1)^{\beta p^{e-1}}g(x))=\omega_H((x-1)^{\beta p^{e-1}}g_1(x))+\omega_H((x-1)^{\beta p^{e-1}}g_2(x)).
\end{align*}

If $g_2(x)\neq0$, from Proposition 2.1, we have $\omega_H(c(x))\geq 2(\beta+1)$. If $\omega_H(c(x))\geq 2\beta+3$, then $\omega_p(c(x))\geq 2\beta+5$. If $\omega_H(c(x))=2\beta+2$, then
\begin{align*}
\omega_H((x-1)^{\beta p^{e-1}}g_1(x))=\omega_H((x-1)^{\beta p^{e-1}}g_2(x))=\beta+1.
\end{align*}

Hence, $\omega_p(c(x))\geq \omega_H(c(x))+\beta+1 =3\beta+3$.

If $g_2(x)=0$, then $g(x)=(g'_1(x))^{p^{e-1}}$, where $g'_1(x)=\sum_{j=1}^sg'_{i_j}x^{i_j}$, $g'^{p^{e-1}}_{i_j}=g_{i_j}$. since $(x-1)\mid g(x)$, so $(x-1)|g'_1(x)$, it can be deduced $c(x)\in \langle (x-1)^{(\beta+1)p^{e-1}}\rangle$, from Proposition 2.1, we have $\omega_H(c(x))\geq \beta+2$. On the other hand, $c(x)$ have the form
\begin{align*}
(c_0~0\cdots 0 ~c_{p^{e-1}}~ 0\cdots 0 ~c_{2p^{e-1}} ~0\cdots 0~ c_{(p-1)p^{e-1}}~0\cdots 0),
\end{align*}
hence, $\omega_p(x)=2\omega_H(c(x))\geq 2(\beta+2)$.

In summary, we have $d_p(\mathscr{C}_{\beta p^{e-1}+1})\geq 2(\beta+2)$. But $\omega_p((x-1)^{(\beta+1) p^{e-1}})=2(\beta+2)$, hence, $d_p(\mathscr{C}_{\beta p^{e-1}+1})=2(\beta+2)$. \qed \\

\noindent \textbf{Proposition 3.5.} \emph{Let $\beta , \gamma, e$ be integer such that $1\leq \beta\leq p-2$, $e\geq 2$ and $ \beta p^{e-1}+1\leq \gamma \leq (\beta+1)p^{e-1}$, then $d_p(\mathscr{C}_\gamma)=2(\beta+2)$.}\\

\textbf{Proof.} From Proposition 3.4, we have $d_p(\mathscr{C}_{\beta p^{e-1}+1})=2(\beta+2)$, for $1\leq \beta \leq p-2$. But
\begin{align*}
(x-1)^{(\beta+1)p^{e-1}}=\sum_{j=0}^{\beta+1}(-1)^{\beta+1-j}{\beta+1 \choose j}x^{jp^{e-1}}.
\end{align*}
Hence, $\omega_p((x-1)^{(\beta+1) p^{e-1}})=2(\beta+2)$. So, $d_p(\mathscr{C}_{(\beta+1) p^{e-1}})\leq 2(\beta+2)$.

For $\beta p^{e-1}+1\leq \gamma \leq (\beta+1)p^{e-1}$, from $\mathscr{C}_{(\beta+1) p^{e-1}}\subseteq \mathscr{C}_\gamma \subseteq \mathscr{C}_{\beta p^{e-1}+1}$, we have
\begin{align*}
2(\beta+2)\leq d_p(\mathscr{C}_{(\beta) p^{e-1}+1}) \leq d_p(\mathscr{C}_\gamma)\leq d_p(\mathscr{C}_{(\beta+1) p^{e-1}})\leq 2(\beta+2),
\end{align*}
Hence, $d_p(\mathscr{C}_\gamma)=2(\beta+2)$. \qed\\

\noindent \textbf{Proposition 3.6.} \emph{Let $k$ be integer such that $1\leq k\leq e-1$, then $d_p(\mathscr{C}_{p^e-p^{e-k}})=2p^k$.}\\

\textbf{Proof.} For any fixed $1\leq k \leq e-1$, let $c(x)$ be any nonzero element of $\mathscr{C}_{p^e-p^{e-k}}$, then there is a nonzero element $a(x)$ and $deg(a(x))\leq p^{e-k}-1$ such that $c(x)=(x-1)^{p^e-p^{e-k}}a(x)$. Then
\begin{align*}
c(x)=\sum_{j=0}^{p^k-1}x^{jp^{e-k}}a(x).
\end{align*}

If $deg(a(x))<p^{e-k}-1$ or $a(0)=0$, then $\omega_p(c(x))=p^k\omega_p(a(x))\geq 2p^k$, and $\omega_p((x-1)^{p^e-p^{e-k}})=2p^k$.

If $deg(a(x))=p^{e-k}-1$ and $a(0)\neq 0$, then $d_p(c(x))=p^k(d_p(a(x))-1)$. If $\omega_H(a(x))\geq 3$, then $\omega_p(a(x))\geq \omega_H(a(x))+1\geq 4$, which deduce that $\omega_p(c(x))\geq 3p^k$. If $\omega_H(a(x))=2$, then $\omega_p(a(x))=4$, which deduce that $\omega_p(c(x))\geq 3p^k$. Hence, $d_p(\mathscr{C}_{p^e-p^{e-k}})=2p^k$. \qed\\

\noindent \textbf{Proposition 3.7.}\emph{ Let $k$ be integer such that $1\leq k\leq e-2$, then $d_p(\mathscr{C}_{p^e-p^{e-k}+1})=3p^k$.}\\

\textbf{Proof.} For any fixed $1\leq k \leq e-2$, let $c(x)$ be any nonzero element of $\mathscr{C}_{p^e-p^{e-k}+1}$, then there is a nonzero element $a(x)$ and $deg(a(x))\leq p^{e-k}-2$ such that $c(x)=(x-1)^{p^e-p^{e-k}}(x-1)a(x)$. Let $b(x)=(x-1)a(x)$, then
\begin{align*}
c(x)=\sum_{j=0}^{p^k-1}x^{jp^{e-k}}b(x).
\end{align*}

If $deg(b(x))<p^{e-k}-1$ or $b(0)=0$, then $d_p(c(x))=p^kd_p(b(x))$, from Proposition 3.1, we have $d_p(c(x))\geq 3p^k$.

If $deg(b(x))=p^{e-k}-1$ and $b(0)\neq 0$, then $d_p(c(x))=p^k(d_p(b(x))-1)$. If $\omega_H(b(x))\geq 3$, then $d_p(b(x))\geq \omega_H(b(x))+1\geq 4$, which deduce that $d_p(c(x))\geq 3p^k$. If $\omega_H(b(x))=2$, then $d_p(b(x))=4$, which means that $d_p(c(x))= 3p^k$. Hence, $d_p(\mathscr{C}_{p^e-p^{e-k}+1})=3p^k$. \qed\\

\noindent \textbf{Proposition 3.8.}\emph{ Let $k, \gamma$ be integer such that $1\leq k\leq e-2$ and $ p^e-p^{e-k}+2 \leq \gamma \leq p^e-p^{e-k}+p^{e-k-1} $, then $d_p(\mathscr{C}_\gamma)=4p^k$.}\\

\textbf{Proof.} For any fixed $1\leq k \leq e-2$, let $c(x)$ be any nonzero element of $\mathscr{C}_{p^e-p^{e-k}+2}$, then there is a nonzero element $a(x)$ and $deg(a(x))\leq p^{e-k}-3$ such that $c(x)=(x-1)^{p^e-p^{e-k}}(x-1)^2a(x)$. Let $b(x)=(x-1)^2a(x)$, then
\begin{align*}
c(x)=\sum_{j=0}^{p^k-1}x^{jp^{e-k}}b(x).
\end{align*}

If $deg(b(x))<p^{e-k}-1$ or $b(0)=0$, then $\omega_p(c(x))=p^k\omega_p(b(x))$, from Proposition 3.1, we have $\omega_p(c(x))\geq 4p^k$.

If $deg(b(x))=p^{e-k}-1$ and $b(0)\neq 0$, then $\omega_p(c(x))=p^k(d_p(b(x))-1)$. If $\omega_H(b(x))\geq 4$, then $\omega_p(b(x))\geq \omega_H(b(x))+1\geq 5$, which deduce that $\omega_p(c(x))\geq 4p^k$. If $\omega_H(b(x))=2$, then $b(x)=x^{p^{e-k}-1}-1$, but $(x^{p^{e-k}-1}-1)\notin \langle (x-1)^2\rangle$, this is contradictory. If $\omega_H(b(x))=3$, then $\omega_p(b(x))\geq 5$. Hence, $\omega_p(c(x))\geq 4p^k$. So, $d_p(\mathscr{C}_{p^e-p^{e-k}+2})\geq 4p^k$. On the other hand,
\begin{align*}
(x-1)^{p^e-p^{e-k}+p^{e-k-1}}=\sum_{j=0}^{p^k-1}x^{jp^{e-k}+p^{e-k-1}}-\sum_{j=0}^{p^k-1}x^{jp^{e-k}}.
 \end{align*}
Hence, $\omega_p((x-1)^{p^e-p^{e-k}+p^{e-k-1}})=4p^k$. So, $d_p(\mathscr{C}_{p^e-p^{e-k}+p^{e-k-1}})\leq 4p^k$.

Let $\gamma$ be integer such that $ p^e-p^{e-k}+2 \leq \gamma \leq p^e-p^{e-k}+p^{e-k-1} $, then
\begin{align*}
4p^k\leq d_p( \mathscr{C}_{p^e-p^{e-k}+2})\leq d_p(\mathscr{C}_{\gamma}) \leq d_p( \mathscr{C}_{p^e-p^{e-k}+p^{e-k-1}})\leq 4p^k.
\end{align*}
Hence, $d_p(\mathscr{C}_{\gamma})=4p^k$.
\qed\\

\noindent\textbf{Proposition 3.9.} \emph{Let $k, \beta, \gamma$ be integer such that $1\leq k\leq e-2$ and $1\leq \beta \leq p-2$, $ p^e-p^{e-k}+\beta p^{e-k-1}+1 \leq \gamma \leq p^e-p^{e-k}+(\beta+1) p^{e-k-1} $, then $d_p(\mathscr{C}_\gamma)=2(\beta+2)p^k$.}\\

\textbf{Proof.} For any fixed $1\leq k \leq e-2$, let $c(x)$ be any nonzero element of $\mathscr{C}_{p^e-p^{e-k}+\beta p^{e-k-1}+1}$, then there is a nonzero element $a(x)$ and $deg(a(x))\leq (p-\beta)p^{e-k-1}-2$ such that $c(x)=(x-1)^{p^e-p^{e-k}}(x-1)^{\beta p^{e-k-1}+1}a(x)$. Let $b(x)=(x-1)^{\beta p^{e-k-1}+1}a(x)$, and $e'=e-k\geq2$, from Proposition 3.4, we have $\omega_p(b(x))\geq 2(\beta+2)$(this can be obtained from the case $e=e'$ in proposition 3.4), then
\begin{align*}
c(x)=\sum_{j=0}^{p^k-1}x^{jp^{e-k}}b(x).
\end{align*}

If $deg(b(x))<p^{e-k}-1$ or $b(0)=0$, then $\omega_p(c(x))=p^k\omega_p(b(x))\geq 2(\beta+2)p^k$.

If $deg(b(x))=p^{e-k}-1$ and $b(0)\neq 0$, then $\omega_p(c(x))=p^k(\omega_p(b(x))-1)$. From the proof of proposition 3.4, we have $d_p(b(x))\geq 2\beta+5$ in this case. Hence, $\omega_p(c(x))\geq 3(\beta+1)p^k\geq 2(\beta+2)p^k$. So, $d_p(\mathscr{C}_{p^e-p^{e-k}+\beta p^{e-k-1}+1})\geq 2(\beta+2)p^k$. Not that $\omega_p((x-1)^{p^e-p^{e-k}+(\beta+1) p^{e-k-1}})=2(\beta+2)p^k$. Hence, for $p^e-p^{e-k}+\beta p^{e-k-1}+1 \leq \gamma \leq p^e-p^{e-k}+(\beta+1) p^{e-k-1}$, we have
\begin{align*}
2(\beta+2)p^k\leq d_p( \mathscr{C}_{p^e-p^{e-k}+\beta p^{e-k-1}+1})\leq d_p(\mathscr{C}_{\gamma}) \leq d_p( \mathscr{C}_{p^e-p^{e-k}+(\beta+1)p^{e-k-1}})\leq 2(\beta+2)p^k.
\end{align*}\qed\\

We summarize the results obtained above in the following theorem.\\

\noindent \textbf{Theorem 3.10.} \emph{Let $\mathcal{C}$ be a cyclic code of length $p^e$ over $\mathbb{F}_{p^m}$, then $\mathcal{C}=\langle (x-1)^i\rangle$, for $i\in \{0,1,\cdots,p^e\}$. The pair-distance $d_p$ of $\mathcal{C}$ is determined by\\
(1) If $e=1$, then
\begin{align*}d_p=
\begin{cases}0& if~i=p,\\
i+2& if ~0\leq i\leq p-2,\\
p& if~i=p-1.
\end{cases}
\end{align*}
(2) If $e\geq2$, then
\begin{align*}d_p=
\begin{cases}0& if~i=p^e,\\
2& if~i=0,\\
3& if~i=1,\\
4& if ~2\leq i\leq p^{e-1},\\
2(\beta+2)& if~\beta p^{e-1}+1\leq i\leq (\beta+1) p^{e-1}, ~where~1\leq \beta\leq p-2,\\
3p^k& if~ i=p^e-p^{e-k}+1, ~where~1\leq k\leq e-2,\\
4p^k& if ~p^e-p^{e-k}+2\leq i\leq p^e-p^{e-k}+p^{e-k-1}, ~where~1\leq k\leq e-2,\\
2(\beta+2)p^k& if~p^e-p^{e-k}+\beta p^{e-k-1}+1 \leq i \leq p^e-p^{e-k}+(\beta+1) p^{e-k-1},\\
& where~1\leq k\leq e-2,~and~1\leq \beta \leq p-2,\\
(j+2)p^{e-1}& if ~i=p^e-p+j,~where~0\leq j\leq p-2,\\
p^e& if ~i=p^e-1.
\end{cases}
\end{align*}}\\

\noindent \textbf{Corollary 3.11.}\emph{ Let $\mathcal{C}$ be a cyclic codes of length $2^e$ over $\mathbb{F}_{2^m}$, then $\mathcal{C}=\langle (x-1)^i\rangle$, for $i\in\{0,1,\cdots,2^e\}$. The pair-distance $d_p$ of $\mathcal{C}$ is determined by \\
(1) If $e=1$, then
\begin{align*}
d_p=\begin{cases}0& if~i=2,\\
2& if~i\in\{0,1\}.
\end{cases}
\end{align*}
(2) If $e\geq 2$, then
\begin{align*}d_p=
\begin{cases}0& if~i=2^e,\\
2& if~i=0,\\
3& if~i=1,\\
4& if ~2\leq i\leq 2^{e-1},\\
3\cdot2^k& if~ i=2^e-2^{e-k}+1, ~where~1\leq k\leq e-2,\\
2^{k+2}& if~ 2^e-2^{e-k}+2\leq i\leq 2^e-2^{e-k}+2^{e-k-1}, ~where~1\leq k\leq e-2,\\
2^e& if~ i=2^e-1.
\end{cases}
\end{align*}}

\dse{4~~Conclusion}
In this paper, the symbol-pair distances of all cyclic codes of length $p^e$ over $\mathbb{F}_{p^m}$ are completely determined. A further research in this area is to construct some new MDS symbol-pair codes via the repeated-root cyclic codes.

\end{document}